\documentclass{PoS}

\PoS{PoS(BDMH2004)020}
\usepackage{epsfig}

\def\simgt{\hbox{\rlap{\raise 0.425ex\hbox{$>$}}\lower 0.65ex\hbox{$\sim$}}}
\def\simlt{\hbox{\rlap{\raise 0.425ex\hbox{$<$}}\lower 0.65ex\hbox{$\sim$}}}
\def\tth{{2/3}}

\title{The phase-space density distribution of dark matter halos}

\ShortTitle{Phase-space density}

\author{\speaker{Liliya L. R. Williams}\\%\thanks{A footnote may follow.}\\
        Univ. of Minnesota, USA\\
        E-mail: \email{llrw@astro.umn.edu}}

\author{Crystal Austin\\
        Univ. of Minnesota, USA\\
        E-mail: \email{caustin@astro.umn.edu}}

\author{Eric Barnes\\
        Univ. of Minnesota, USA\\
        E-mail: \email{barnes@astro.umn.edu}}

\author{Arif Babul\\
        Univ. of Victoria, B.C., Canada\\
        E-mail: \email{babul@uvic.ca}}

\author{Julianne Dalcanton\\
        Univ. of Washington, USA\\
        E-mail: \email{jd@astro.washington.edu}}

\abstract{
High resolution N-body simulations have all but converged on a common  
empirical form for the shape of the density profiles of halos,
but the full understanding of the underlying physics of halo
formation has eluded them so far. We investigate the formation
and structure of dark matter halos using 
analytical and semi-analytical techniques. Our halos are formed
via an extended secondary infall model (ESIM); they contain
secondary perturbations and hence random tangential and
radial motions which affect the halo's evolution at it undergoes
shell-crossing and virialization.
Even though the density profiles of NFW and ESIM  halos are
different their phase-space density distributions
are the same: $\rho/\sigma^3\propto r^{-\alpha}$, with $\alpha=1.875$ 
over $\sim 3$ decades in radius. We use two approaches
to try to explain this ``universal'' slope: (1) The Jeans equation
analysis yields many insights, however, does not answer why
$\alpha=1.875$. (2) The secondary infall model of the 1960's and 
1970's, augmented by ``thermal motions'' of particles does 
{\em predict\/} that halos should have $\alpha=1.875$. However,
this relies on assumptions of spherical symmetry and slow accretion.
While for ESIM halos these assumptions are justified, 
they most certainly break down for simulated 
halos which forms hierarchically. We speculate that our argument
may apply to an ``on-average'' formation scenario of halos within
merger-driven numerical simulations, and thereby explain why
$\alpha=1.875$ for NFW halos. ~Thus, $\rho/\sigma^3\propto r^{-1.875}$
may be a generic feature of violent relaxation.
}

\FullConference{Baryons in Dark Matter Halos\\
		 5-9 October 2004\\
		 Novigrad, Croatia}

\begin{document}

\section{Introduction}

There is a broad consensus that gravity driven evolution of the 
space distribution of collisionless dark matter, starting from 
some realistic matter power spectrum results in virialized objects
whose spherically averaged density profile is well represented by the
NFW prescription \cite{nfw97} or its close variants \cite{m98}. 
However, why the dark matter profiles have this shape is as yet
to be determined. 
In an effort to shed light on the issue, \cite{tn01} investigated 
the phase-space structure of dark matter halos from N-body simulations.
They found that the ``poor man's'' phase-space density, 
$f(r)=\rho/\sigma^3$, is a power law over 3 decades in radius, 
$f(r)\propto r^{-\alpha}$, with exponent $\alpha=1.875=15/8$.

We have developed an alternative method of generating dark matter halos,
based on \cite{rg87}. It is an analytical scheme which treats collapse,
shell-crossing and virialization of spherically symmetric halos. The
halo contains secondary perturbations whose properties are calculated 
from the same power spectrum that gives rise to the main halo. These 
secondary perturbations induce random tangential and radial motions 
within the halo. The ensuing collapse can be likened to slow accretion 
of lumpy matter; there are no major mergers in our scheme; \cite{wbd04}. 
We call these halos ESIM, or Extended Secondary Infall Model halos. 
Even though ESIM and NFW halos are generated in very different ways, and 
have different density profiles, their phase-space density distribution, 
and the value of $\alpha$ is the same for both. Our goal
is to understand why is $\alpha=1.875=15/8$ for both.

\section{The Jeans equation analysis of phase-space density distribution}
\label{jeans}

Equilibrium, non-rotating halos with isotropic
velocity ellipsoids obey this Jeans equation:
\begin{equation}
{d\over{dr}}\Bigl[{{-r^2}\over{G\rho}}{{d(\rho\sigma^2)}\over{dr}}\Bigr]
={d\over{dr}}M(<r).
\label{tn01eq3}
\end{equation}
Following \cite{tn01} we use scaled variables, $x=r/r_0$, and $y=\rho/\rho_0$.
We assume power law behavior of $f(r)$ with constant $\alpha$ for any 
given halo, but allow the density profile to have changing log-log slope,
$y\propto x^{-\beta(x)}$. With these,
eq.~\ref{tn01eq3} can be rewritten in terms of $x$, $y$, 
exponents $\alpha$ and $\beta$, and a normalization constant. 
Unlike \cite{tn01} we work with an equation obtained by differentiating 
eq.~\ref{tn01eq3}:
\begin{equation}
15\beta^{\prime\prime}-3\beta^\prime(8\alpha-5\beta-5)=
(2\alpha+\beta-6)(2\alpha-5\beta)(\tth[\alpha-\beta]+1)
\label{b1}
\end{equation}
Here, $\beta^\prime$ and $\beta^{\prime\prime}$ are derivatives of 
$\beta$ with respect to $-\ln x$. (Recall that $\beta=-d\ln y/d\ln x$.) 

Because eq.~\ref{b1} has many solutions (i.e. halo density profiles), 
it would help to make analytic inroads into the analysis of its solution 
space. We do just that: we derive a family of solutions:
\begin{equation}
\beta^\prime=\beta^\prime_m-\tth(\beta-\beta_0)^2,\quad
\beta^\prime_m=3(2\alpha+5)^2/200,\quad
\beta_0=(14\alpha+15)/20.
\label{bpX}
\end{equation}  
All the halos that obey these three equations asymptote to very nearly 
constant $\beta$ slopes at large and small radii, and have non-zero derivatives 
of $\beta$ in-between. We will refer to this as the {\em main family} of 
solutions. Each member is completely defined by specifying $\alpha$. As far as 
we can tell, this is the only family that has a closed form analytical 
description. Note that the `critical solution' identified by \cite{tn01} 
is a member of this family; it is obtained by setting $\alpha=1.875$.  

\begin{figure}
\epsfig{file=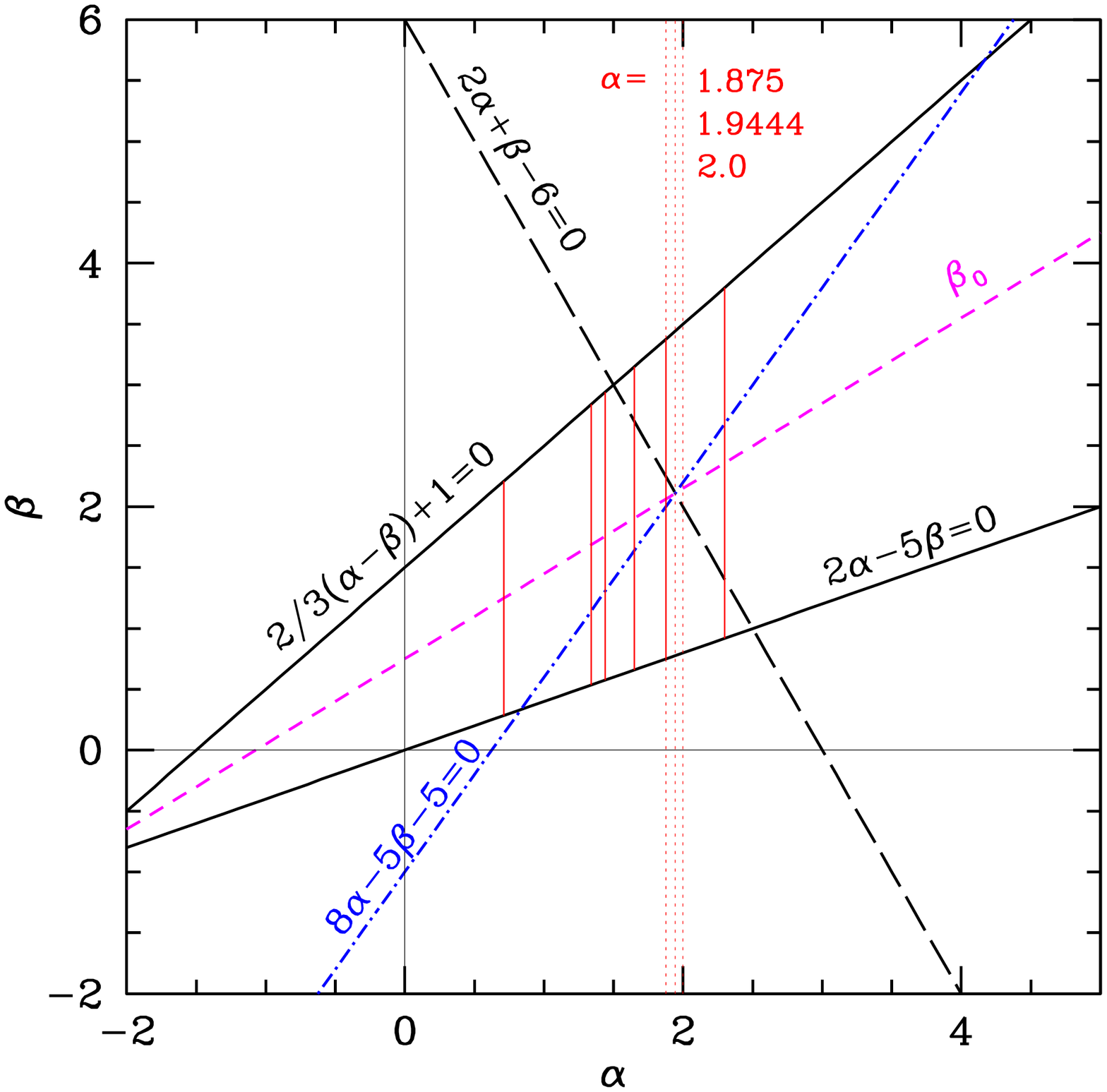, width=0.49\textwidth}
\epsfig{file=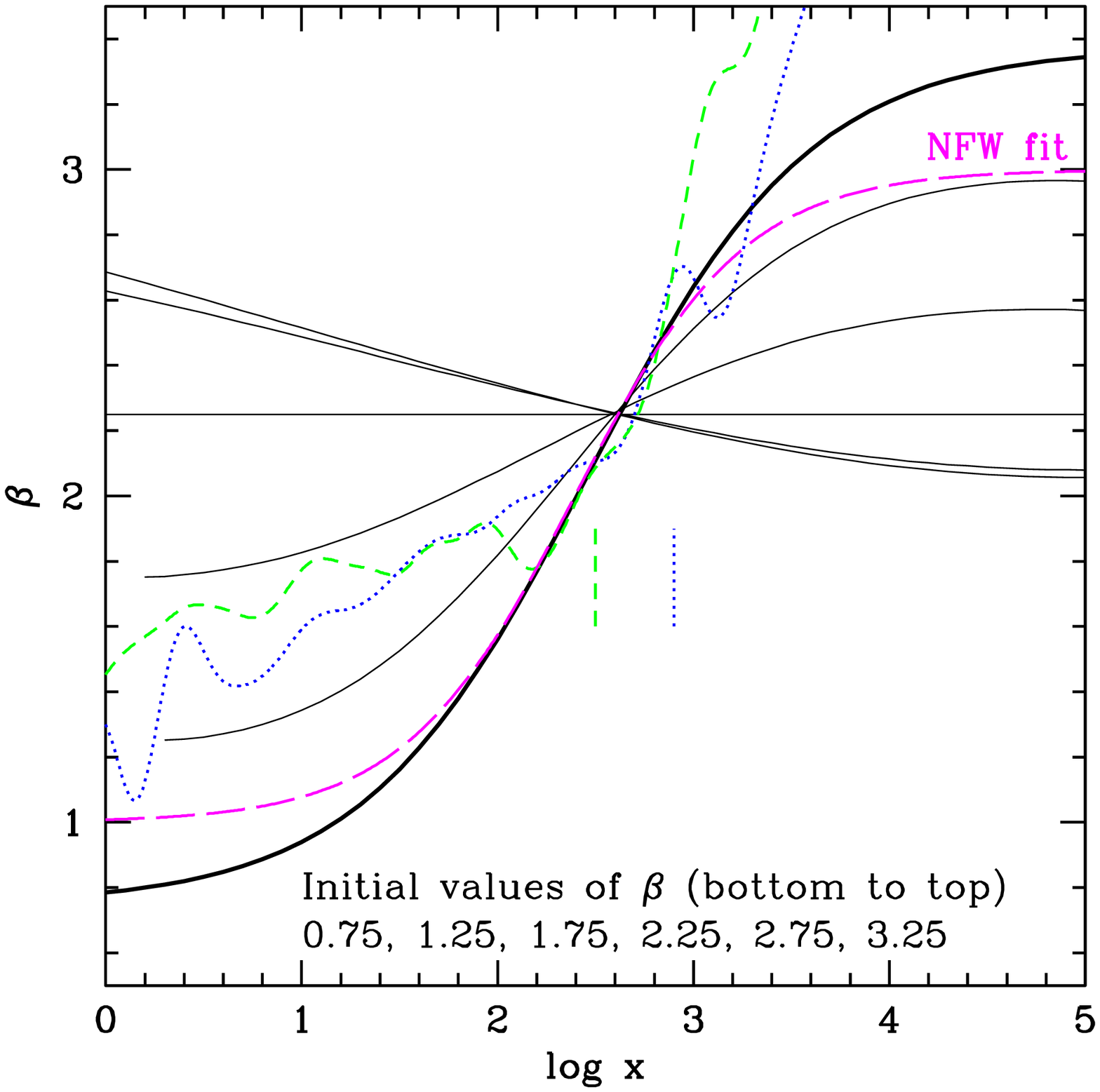, width=0.49\textwidth}
\caption{
{\it (a)~} 
Pictorial depiction of solutions of eq.~2.2. $\alpha=1.875$ is
the slope of the phase-space density profiles of NFW and ESIM halos; 
$\alpha=1.9444$ results in periodic, zero-damping solutions of eq.~2.2.
Six solid vertical line segments are examples of the main family of solutions.
Labeled lines: factors in eq.~2.2; $\beta_0$ is given in eq.~2.3.
{\it (b)~}
A set of solutions of eq.~2.2, with constant $\alpha=1.875$ but varying 
initial values of $\beta=1.25,1.75,...3.25$; initial $\beta^\prime=0$ for all. 
The thick line belongs to the main family of solutions; note that NFW
empirical fit follows this solution closely for intermediate values of $\beta$.
Dotted and dashed lines are two ESIM halos; short vertical line segments 
indicate their virial radii. Normalization of radius $x$ is arbitrary.}
\label{smtrifam}
\end{figure}

Figure~\ref{smtrifam}(a) is a schematic representation of all solutions of 
eq.~\ref{b1}. A few main family solutions are shown as solid vertical red 
line segments; the inner density cusp for each
is given by $(2\alpha-5\beta_{in})=0$, the outer density profile 
slope is given by $(\tth[\alpha-\beta_{out}]+1)=0.$ Fig.~\ref{smtrifam}(a)
illustrates the role these factors on the RHS of eq.~\ref{b1} play in defining
main family solutions.
 
One of these six vertical line segments is for $\alpha=1.875$; the corresponding
$\beta(x)$ is shown as thick solid curve in fig.~\ref{smtrifam}(b). Other curves 
are what we call $\alpha$-family solutions, obtained by keeping  $\alpha$
constant, but varying initial conditions: $\beta$ and $\beta^\prime$. 
Note that one of the members of the $\alpha$-family is a power law density 
profile, $\beta$=const. The $\beta$ value for the power law solution of an
arbitrary $\alpha$-family is obtained by solving $(2\alpha+\beta-6)=0$;
see eq.~\ref{b1} and long-dash line in fig.~\ref{smtrifam}(a).

Does eq.~\ref{b1} allow any special $\alpha$ values? The equation is
a non-linear damped oscillator. For some $\alpha$ the $x$-averaged value 
of the 'dissipation' term, $-3\beta^\prime(8\alpha-5\beta-5)$ is zero. 
This value is $\alpha=35/18\approx1.9444$. When $\alpha=35/18$ any 
initial condition for $\beta$ and $\beta^\prime$ results in a periodic, 
constant amplitude function of $\beta$ vs. $x$.
Value $\alpha=35/15$ is only $\sim4\%$ different from $\alpha=1.875$. 
Aside from this coincidence
(or not?), and many other interesting insights provided by the analysis of 
eq.~\ref{b1}, we were not able to identify why $\alpha=1.875$ is special. 
Next, we take a different, and more promising approach to uncover why 
$\alpha=1.875$ for ESIM and NFW halos alike.

\section{Secondary infall analysis of phase-space density distribution}

Consider the initial stages of halo collapse in the context of secondary 
infall models. A small constant central mass excess, $\delta M_0$ is 
surrounded by material of average density. The dynamics of the pre-turn-around 
period is described by the parametric equations of \cite{gg72}.
%$r\propto (1-\cos\theta)$ and $t\propto (\theta-\sin\theta)$. 
The turn-around radius for a shell of initial comoving $r_i$, is 
$r_m=r_i\bar\delta^{-1}$, and $\bar\delta$ is the initial fractional 
overdensity inside $r_i$. Then, $M(<r_i)\propto\rho r_i^3$, so 
$r_i\propto M^{1/3}$. Also, $\bar\delta=\delta M_0/M\propto M^{-1}$.  
Combining these scalings we get $r_m\propto M^{4/3}$. After reaching 
turn-around each shell collapses back a little; one typically 
assumes a constant collapse factor. Assume that each shell spends most 
of the time at its 
apocenter, and so most of its mass is located at that radius, which 
is $\propto r_m$. The resulting density distribution in the proto-halo is,
$\rho(r)\propto r^{-2}\,dM/dr$, or,
$\rho\propto M^{-3}\propto r_m^{-9/4}$. This is a well known result.

In the real Universe the collapse of material will not be purely radial;
there will be some random motion of particles, and associated kinetic
energy. We speculate that during the early stages of collapse the 
kinetic energy will be derived from the gravitational 
potential energy, and therefore will be proportional to 
the potential energy: ${1\over 2} v^2\propto GM/r_m$. Kinetic energy
of random motion gives an estimate of the velocity dispersion: 
$\sigma\propto [M/r_m]^{1/2}\propto r_m^{-1/8}$. 
So, $\sigma^3\propto r_m^{-3/8}$. Combining $\rho(r)$ from the previous
paragraph with $\sigma(r)$, we get 
$\rho/\sigma^3\propto (r^{-9/4})/(r^{-3/8})\propto r^{-15/8}$,
~i.e. $\alpha=1.875$!

The result obtained above says that the phase-space density is a function 
of $r$. Because $E\propto GM/r$, $E$ is a monotonic function of $r$, at
least for systems that are, on average, spherically symmetric.
In equilibrium, the total energy $E$ of a particle is its integral of motion.
If the halo collapse proceeds slowly then the halo passes through a 
series of quasi-equilibrium stages.  Maybe we can assume that energy is an 
`approximate' integral of motion in a slowly collapsing galaxy. In that
case, our relation $\rho/\sigma^3\propto r^{-1.875}$ can be interpreted as
saying that the phase-space density ($\rho/\sigma^3$) is a function of
$E$ only, in compliance with Jeans Theorem.

If dark matter is collisionless, then the collapse will preserve the 
phase-space density as calculated above. So the final virialized dark 
matter halos, whatever their density and velocity dispersion profiles, 
will have the same phase-space density distribution that was characteristic 
of the early stages of collapse. Therefore, virialized halos are expected 
to have $\rho/\sigma^3\propto r^{-1.875}$. This argument relies on many
approximations, most of which can be justified for spherically symmetric,
smoothly accreting ESIM halos. However, one is hard pressed to see why
these approximations would hold in a hierarchical formation model. We 
speculate that the argument could apply to an 'average' situation; 
after all, NFW is an average shape of numerically generated halos.  

Finally, we note the connection between the above argument and the 
$\alpha$-family solutions of \S~\ref{jeans}. The various evolutionary stages
of halos have different density profiles. We claim that all these are 
represented by members of the {\em same\/} $\alpha=1.875$-family. The 
earliest epoch halos have power law density profiles with $\beta=9/4$, 
value derived at the top of this \S; the same value is derived
using Jeans equation analysis---the horizontal line in 
fig.~\ref{smtrifam}(b). Later stages of halo evolution develop changing
slope density profiles, represented by more and more curved lines in
fig.~\ref{smtrifam}(b). The final stage, the thick solid line is a good
approximation to the NFW profile obtained from simulations.


\begin{thebibliography}{99}

%\bibitem{b85}%[Bertschinger(1995)]
%Bertschinger, E. 1985, ApJS, 58, 39 

\bibitem{gg72}
Gunn, J.E. \& Gott, J.R. 1972, ApJ, 176, 1

\bibitem{m98}%[Moore et al.(1998)]
Moore, B., Governato F., Quinn, T., Stadel, J. \& Lake, G. 1998,
ApJ, 499, L5

\bibitem{nfw97}%[Navarro et al.(1997)]
Navarro, J.F., Frenk, C.S. \& White, S.D.M. 1997, ApJ, 490, 493

\bibitem{tn01}%[Taylor \& Navarro(2001)]
Taylor, J.E. \& Navarro, J.F. 2001, ApJ, 563, 483

\bibitem{rg87}
Ryden, B.S. \& Gunn, J.E. 1987, ApJ, 318, 15

\bibitem{wbd04}%[Williams et al.(2004)]
Williams, L.L.R., Babul, A. \& Dalcanton, J.J. 2004, ApJ, 604, 18

\end{thebibliography}
\end{document}